\documentstyle{aipproc}

\newcommand{\beq}{\begin{equation}}
\newcommand{\eeq}{\end{equation}}
\newcommand{\beqa}{\begin{eqnarray}}
\newcommand{\eeqa}{\end{eqnarray}}

\begin{document}

\title{Linear dynamics of the solar convection zone: excitation of
waves in unstably stratified shear flows}
\author{G.~D.~Chagelishvili$^{*,\dagger}$, A.~G.~Tevzadze$^{*,\P}$
M.~Goossens$^{\P}$}
\address{$^*$ Abastumani Astrophysical Observatory, 2$^a$ Ave. Kazbegi 380060
Tbilisi, Georgia \\
$^\dagger$ Space Research Institute, 84/32 Str. Profsoyuznaya, 117810
Moscow, Russia \\
$^\P$ Center for Plasma Astrophysics, K.U.Leuven, Celestijnenlaan 200B,
3001, Heverlee, Belgium}
\maketitle

\begin{abstract}

In this paper we report on the nonresonant conversion of
convectively unstable linear gravity modes into acoustic
oscillation modes in shear flows. The convectively unstable linear
gravity modes can excite acoustic modes with similar wave-numbers.
The frequencies of the excited oscillations may be qualitatively
higher than the temporal variation scales of the source flow,
while the frequency spectra of the generated oscillations should be
intrinsically correlated to the velocity field of the source flow.
We anticipate that this nonresonant phenomenon can significantly
contribute to the production of sound waves in the solar
convection zone.

\end{abstract}

\section*{INTRODUCTION}

The excitation and propagation of waves are important for
understanding the dynamics of the sun and stars.
It is believed that most of the solar mechanical energy is
accumulated in the turbulent motions in its convection zone.
In the convection zone the gravitational stratification drives
the convective instability providing the dynamical activity of
this relatively thin region. The dynamics of the solar convection
is studied to explain many observational features of the Sun.
Notably, it is thought that the solar acoustic oscillations are
excited by the turbulence in the convection zone [1-5].

Lighthill's ideas of aerodynamic sound generation form the basis
of the theoretical investigation of the wave excitation in a
hydrodynamic medium [6,7].
This theory of wave excitation by a free turbulence has
been generalized for stratified fluids by Stein [8].
From a physical point of view, Lighthill's theory of wave
generation employs the concept of stochastic excitation of
oscillations (waves). In Lighthill's theory perturbations
are described by an inhomogeneous wave equation, with linear
terms forming the oscillatory part and the inhomogeneous terms
standing for the source function. The source terms, which may
be classified by their multipole order, are stochastically
created by the turbulent perturbations.
The amplifying effect of a sheared mean flow on
the fluctuations of the Reynolds stress (nonlinear source term)
and thus on the wave production has been noted by Lighthill [7].
However, this effect has not received further attention within the
context of stochastic excitation.

Significant advances in the investigation of the dynamics of flows
with velocity shear have been achieved together with the
disclosure of specific features of shear flow phenomena
[9,10]. Operators arising in the mathematical
formalism of the canonical modal analysis in the study of the
linear dynamics of shear flows are not self-adjoint. Consequently
eigenmode interference introduces principal complications.
The nonmodal approach has proved to be an alternative
successful route for exploring the dynamics of shear flows. This
approach employs the study of temporal evolution of the spatial
Fourier harmonics of perturbations.

Impressive progress has been made by use of the nonmodal analysis
(see e.g., [11-16]). This approach has led to the discovery of
new channels of energy exchange between different modes in shear
flows. Resonant phenomena of wave transformations have been
studied in [17-23]. The nonresonant phenomenon of the conversion
of vortices into acoustic waves has been described in [24]. The
same mechanism is found to operate for magnetosonic [25] as well
as for plasma Langmuir oscillations [21].

In this report we introduce a new dynamical source of acoustic
waves in unstably stratified shear flows. Namely, the {\it linear}
nonresonant conversion of convective into acoustic wave modes in a
stratified shear flows.
Convectively unstable exponentially growing buoyancy perturbations
generate acoustic wave oscillations in presence of a sheared mean
flow. We identify this linear conversion of modes in shear flows
as a new excitation mechanism of the solar oscillations and
waves. It differs in principle from the stochastic excitation
mechanism and should significantly contribute to the process of
acoustic wave generation in the solar convection zone.

\section*{Physical approach}

The equations governing the dynamics of a compressible stratified
flow are:
$$
\left[\partial_t + ({\bf V \nabla}) \right]
 \rho + \rho ({\bf \nabla V}) = 0, \eqno(1.a)
$$
$$
\left[ \partial_t + ({\bf V \nabla}) \right] {\bf V} =
-{{\bf \nabla} P / \rho} + {\bf g}, \eqno(1.b)
$$
$$
\left[
\partial_t + ({\bf V \nabla}) \right] P = (\gamma P / \rho)
\left[ \partial_t + ({\bf V \nabla}) \right] \rho. \eqno(1.c)
$$
We consider the hydrodynamic situation where a horizontal shear
flow $\mbox{\bf V}_0=(Ay,0,0)$ occurs in a vertically
stratified  medium $\mbox{\bf g}=(0,0,-g)$.
For simplicity we assume that $A=const$ and $g=const$.
This yields the stratified equilibrium state:
$$
{P_0(z) / P_0(0)} = {\rho_0(z) / \rho_0(0)} = \exp(-zk_H),
\eqno(2)
$$
where $k_H \equiv \gamma g / c_s^2 $ and $c_s^2 \equiv \gamma P_0
/ \rho_0$. We introduce the linear perturbations in the
following way:
$$
\mbox{\bf V} = \mbox{\bf V}_0 + \mbox{\bf V}^\prime \rho_0(0) /
\rho_0(z),  ~~ P=P_0+P^\prime, ~~ {\rho} =
\rho_0 + \rho^\prime. \eqno(3)
$$
Here the velocity perturbations
are normalized to exclude the exponential height dependence due to
the vertical stratification of the background flow. We use the
Cowling approximation [26] and neglect the perturbations of
the gravitational acceleration. Following the standard method of
nonmodal analysis (see [27] for a rigorous mathematical
interpretation) we introduce the spatial Fourier harmonics (SFH)
of the perturbations with time dependent phases:
$$
\Psi(\mbox{\bf r},t) = \psi(\mbox{\bf k}(t),t) \exp ( {\rm i}k_xx
+ {\rm i}k_y(t)y + {\rm i}\tilde k_zz), \eqno(4.a)
$$
$$
k_y(t) = k_y(0)-Ak_xt, \eqno(4.b)
$$
where $\tilde k_z \equiv k_z + {\rm i} k_H/2$. For compactness
of notation we introduce the generalized vector of perturbations
and their SFHs as follows: $ \Psi \equiv (\mbox{\bf V}^\prime,
p^\prime, \rho^\prime)$ and $\psi \equiv (\mbox{\bf u}, p, \rho)$.
To avoid complex coefficients in the dynamical equations, we
construct the normalized entropy and vertical velocity
perturbation SFHs in the following way:
$$
s \equiv ({\rm i}c_s^2 \tilde k_z^*/g - 1 )
(p-c_s^2 \rho) /(\gamma - 1), \eqno(5.a)
$$
$$
v \equiv (c_s^2 \tilde k_z^* + {\rm i}g )u_z, \eqno(5.b)
$$
where $\tilde k_z^*=k_z-{\rm i}k_H/2$. From Eqs. (1-5)
we obtain, by the use of straightforward manipulations, the
following set of differential equations that govern the SFH of the
linear perturbations in stratified shear flow:
$$ {\buildrel
\mbox{\large \bf .} \over p}(t) = c_s^2 (k_x u_x  + k_y(t) u_y) +
v, \eqno(6.a)
$$
$$
{\buildrel \mbox{\large \bf .} \over u_x}(t) =
-A u_y - k_x p , \eqno(6.b)
$$
$$
{\buildrel \mbox{\large \bf .}
\over u_y}(t) = -k_y(t) p , \eqno(6.c)
$$
$$
{\buildrel
\mbox{\large \bf .} \over v}(t) = (N_{\rm B}^2 - c_s^2 \bar k_z^2)
p - N_{\rm B}^2 s, \eqno(6.d)
$$
$$
{\buildrel \mbox{\large \bf .}
\over s}(t) = v. \eqno(6.e)
$$
$N_{\rm B}^2$ is the square of the
frequency of the Brunt-V\"ais\"al\"a: $ N_{\rm B}^2 \equiv gk_H
(\gamma - 1) / \gamma$ and $\bar k_z^2 = |\tilde k_z|^2 =
k_z^2+k_H^2/4$. In an unstably stratified flow negative buoyancy
($N_{\rm B}^2<0$) requires that the adiabatic index $\gamma < 1$.
Such an effective value may be assigned to this parameter under a
certain thermodynamic approach [28]. However, in Eqs
(6.a-e) we retain only $N_{\rm B}^2$ and argue that these
equations are more general than the underling $\gamma$
prescription.

Further we note that vorticity is conserved in the wave-number
space: $I = k_x u_y - k_y(t) u_x - (A /c_s^2)(p - s)$.
The spectral energy of the perturbations can be defined as follows:
$$
E = {\rho_0 /(2 c_s^2}) \left( E_{\rm K} + E_{\rm P} +
E_{\rm T} \right), \eqno(7.a)
$$
$$
E_{\rm K} = c_s^2 ( u_x^2 + u_y^2 ) +
v^2 / {(c_s^2 \bar k_z^2 - N_{\rm B}^2)}, \eqno(7.b)
$$
$$
E_{\rm P} = p^2,
~~~~
E_{\rm T} = {N_{\rm B}^2 s^2 / (c_s^2 \bar k_z^2 - N_{\rm B}^2)}.
\eqno(7.c,d)
$$
where $E_{\rm K}$, $E_{\rm P}$ and $E_{\rm T}$
correspond to the kinetic, elastic and thermobaric energies of the
perturbations, respectively. Formally the perturbation energy is
conserved in the shearless limit: ${\buildrel \mbox{\large \bf .}
\over E} = A c_s^2 u_x u_y$. The instability of the convective
eddies corresponds to a negative value of the thermobaric energy.

\subsection*{Linear modes in the shearless limit}

The linear modes may be classified explicitly in the shearless
limit ($A=0$). In this case the full Fourier expansion of the
linear perturbations
$\Psi({\rm r},t) \propto \tilde \psi({\rm k},\omega)$
yields the dispersion equation:
$$
\omega(\omega^4 - c_s^2 k^2 \omega^2 + N_{\rm B}^2 c_s^2
k_\perp^2) = 0, \eqno(8)
$$
where $k_\perp^2 \equiv k_x^2+k_y^2$ and $k^2 = k_\perp^2 + \bar k_z^2$.
The solutions of Eq. (8) describe the stability and characteristic
temporal variation scales of the existing modes:
$$
\omega_{\rm v}=0, \eqno(9.a)
$$
$$
\omega_{s,c}^2 = {1 \over 2} c_s^2 k^2 \left\{1 \pm \left({1 -
{4 N_{\rm B}^2 k_\perp^2 \over c_s^2 \bar k^4} } \right)^{1/2}
\right\} , \eqno(9.b)
$$
where the subscripts v, {\it s, c} define the frequencies of the
vortex, acoustic and convective modes, respectively.
In an unstably stratified flow, i. e., when $N_{\rm B}^2<0$,
i$\omega_c$ defines the growth rate of the buoyancy
perturbations.

Obviously the $I=constant$ law demonstrates the existence of the
stationary ($\omega=0$) vortex mode in the linear spectum.
The conserved vorticity $I$ may be considered as the vortex mode
measure. The physical eigenfunctions of the acoustic $\Phi_s(t)$
and convective $\Phi_c(t)$ modes may be rigorously defined
in this limit:
$$
\Phi_s(t) \equiv p(t) + N_{\rm B}^2{\Omega_s^2-\omega_s^2 \over
\Omega_c^4} \left( s(t) - {\bar k_z^2 \over k^2} p(t) \right)
\eqno(10.a)
$$
$$
\Phi_c(t) \equiv s(t) - {\bar k_z^2 \over k^2} p(t) -
{{\Omega_c^2-\omega_c^2 \over N_{B}^2}}p(t) \eqno(10.b)
$$
where $\Omega_s^2 \equiv c_s^2 k^2$ and
$ \Omega_c^2 \equiv N_{\rm B}^2 {k_\perp^2 / k^2}$.
Hence the equations governing the dynamics of the perturbations of
the different modes may be decoupled as follows:
$$
{\buildrel \mbox{\large \bf ..} \over \Phi_s}(t) + \omega_s^2
\Phi_s(t) = 0, \eqno(11.a)
$$
$$
{\buildrel \mbox{\large \bf ..} \over \Phi_c}(t) + \omega_c^2
\Phi_c(t) = 0. \eqno(11.b)
$$
Starting from this simple situation we study the velocity shear
effects on the perturbation modes.

\subsection*{Effects of a sheared flow}

To study the effects of the velocity shear on the linear
modes we introduce the small-scale approximation:
$k_z^2 \gg k_H^2$. This approximation strongly simplifies the
mathematical formulation and is justified for the
following two reasons. Firstly, our analysis needs constant
vertical gravity, an assumption that may be adopted for
perturbations with vertical height scales shorter than the
stratification scale. Secondly, this approximation is
necessary for our assumption of a constant linear shear
of the flow velocity, especially in the turbulent flows.
Using Eq. (7) this approximation may be represented by the
following condition: $c_s^2 k_z^2 \gg N_{\rm B}^2$.
In terms of the frequencies it yields
$(\Omega_s^2-\omega_s^2) \approx (\Omega_c^2-\omega_c^2) \approx 0$,
which strongly simplifies the characteristic physical quantities
of the perturbation modes: $\Phi_s(t) \approx p(t)$ and
$\Phi_c(t) \approx (s(t) - \bar k_z^2 p(t)/ k^2(t))$.
To analyze the dynamics of acoustic oscillations in the shear
flow we rewrite Eqs. (6.a-e) in the form of coupled second order
differential equations for the variables $p(t)$ and $y(t)$:
$$
{\buildrel \mbox{\large \bf ..} \over p}(t)
+ f(t){\buildrel \mbox{\large \bf .} \over p}(t)
+ \Omega_1^2(t) p(t) =
\lambda_{1}(t){\buildrel \mbox{\large \bf .} \over y}(t)
+ \lambda_{2}(t) y(t),  \eqno(12.a)
$$
$$
{\buildrel \mbox{\large \bf ..} \over y}(t)
+ \Omega_2^2(t) y(t) = 0, ~~~~~~~~~~ \eqno(12.b)
$$
where the convection variable $y(t)$ is introduced as follows:
$$
y(t) \equiv {k_\perp(t) \over k(t)~} \left( s(t) -
{\bar k_z^2 \over k^2(t)} p(t) \right)
\eqno(13)
$$
and
$$
\Omega_1^2(t) = c_s^2 k^2(t) + 2A^2{k_x^2 \over k^2(t)}
-4A^2 {k_x^2 k_y^2(t) \bar k_z^2 \over k_\perp^2(t) k^4(t)},
\eqno(14.a)
$$
$$
\Omega_2^2(t) = N_{\rm B}^2 {k_\perp^2(t) \over k^2(t)} + 2A^2
{k_x^2k_z^2 \over k_\perp^4(t) k^4(t)}
\left[ 3k_\perp^2(t) k^2(t) -
4 k_y^2(t)k_\perp^2(t)-k_y^2(t) \bar k_z^2 \right],
\eqno(14.b)
$$
$$
f(t)=2A{k_xk_y(t) \over k^2(t)},
\eqno(14.c)
$$
$$
\lambda_{1}(t) = -2A{k_xk_y(t) \over k_\perp(t) k(t)},
\eqno(14.d)
$$
$$
\lambda_{2}(t) = -2A^2 {k_x^2 k_\perp(t) \over k^3(t)}
\left( 1-{k_y^2(t) \bar k_z^2 \over k_\perp^4(t)} \right).
\eqno(14.e)
$$
In deriving Eqs. (12.a-b) we have used the following two
simplifications.
Firstly, we have retained only the
terms describing the effect of the buoyancy perturbations on
the acoustic waves, and we have neglected the effect of the acoustic
pressure perturbations on the evolution (exponential amplification)
of the buoyancy perturbations in the right hand side (rhs)
of Eq. (12.b). Secondly, we have neglected the source terms in the rhs
of the two dynamical equations that describe the shear induced
coupling between the vortex and acoustic wave modes (in Eq. 12.a)
and vortex and buoyancy modes (in Eq. 12.b).
In fact, the coupling of the vortex and acoustic wave
is a process that has
been studied to reveal the mean flow shear induced nonresonant
mode conversion phenomenon in [24]. However, in the present
case, the source terms of the acoustic waves that are proportional to
the vortex mode measure, conserved quantity $I$, are
dominated by the source terms, associated with the exponentially
amplifying convective modes: $y(t)$ and
${\buildrel \mbox{\large \bf .} \over y}(t)$.
It should be emphasized that the present approach is
justified only for a convectively unstable medium with
$N_{\rm B}^2 <0$, so that the buoyancy modes undergo exponential
amplification in the linear regime.

The dynamics of the acoustic waves in the absence of the buoyancy
perturbations is described by the homogeneous part of Eq. (12.a).
The acoustic wave frequency and amplitude variations are described
by the parameters $\Omega_1^2(t)$ and $f(t)$
(see [19] for a detailed study).
The dynamics of the convective mode is described by
Eq. (12.b). Eq. (16.b) shows the transient stabilization effect
of the sheared mean flow in an unstably stratified medium.
The stabilization occurs at times, when $|k_y(t)/k_x| < 1$
and reaches its maximum at $t=t^*$, when $k_y(t^*)=0$ (see Eq. 16.b).

The terms $\lambda_{1}(t){\buildrel \mbox{\large \bf .} \over y}(t)$
and $\lambda_{2}(t) y(t)$ in the rhs of Eq. (14.a)
describe the coupling between the convective and acoustic waves modes.
The shear flow origin of these source terms is obvious from Eqs. (16.d,e).
Hence, Eqs. (14.a,b) describe the mean flow shear induced
buoyancy -- acoustic wave mode conversion in a convectively
unstable medium.
Some specific features of this phenomenon are due to its linear
nature; SFH of the exponentially growing buoyancy perturbations are able
to generate SFH of the acoustic waves with the same wave-numbers.
The amplitude of the excited wave mode depends on the values of the source terms
$\lambda_1(t)$ and $\lambda_2(t)$. So, convective modes with
$k_x=0$ can not generate acoustic waves at all ($\lambda_1=\lambda_2=0$).
While maximal efficiency of the mode conversion phenomenon should
occur at $k_z=0$, or in a realistic physical approximation (see Eq. 12)
at $k_z^2 \geq N_{\rm B}^2 / c_s^2$. Naturally, acoustic wave emission from
convection should generally increase when the mean flow
shear parameter $A$ increases.

We numerically analyze  Eqs. (6.a-e) to verify the analytical
results obtained from the approximate equations (12.a,b). We
select the initial perturbations in a specific manner, which enables
us to excite the convective and acoustic wave modes
individually at the initial moment of time. It appears that
exponentially growing buoyancy perturbations instantly excite the
acoustic wave mode harmonics at a given point in time, when the
perturbation wave-number along the flow velocity shear is zero:
$t=t^*$, $k_y(t^*)=0$. The generation of acoustic waves is clearly
traced from the pressure variation, as well as the
compression of the perturbations. Numerical analysis shows that the
efficiency of this mode conversion phenomenon increases with the
flow shear parameter.

\section*{DISCUSSION AND CONCLUSIONS}

We have presented a study of compressible convection in shear
flows. In particular we have focused on linear small-scale
perturbations in unstably stratified flows with constant shear of
velocity. The linear character of the system enables us to
identify the perturbation modes and to study their dynamics
individually. We find a mode conversion that originates
from the velocity shear of the flow: exponentially growing
perturbations of convection are able to excite acoustic waves.
This process offers a novel approach to the hydrodynamic problem
of the acoustic wave generation.

This wave excitation phenomenon can be
important for the acoustic oscillations of the sun.
Being responsible for the wave generation in high shear
regions of a stratified turbulent flow, this nonresonant
phenomenon can contribute to
the production of sound in the solar convection zone.
Moreover, the process of the wave excitation should be triggered
by a weak vertical magnetic field. In this case we anticipate the
production of high frequency compressional MHD waves. The
latter process will considerably increase the extraction of the
mechanical energy of the convection by waves.

Specific to this phenomenon is that perturbations of buoyancy
are able to excite acoustic waves with similar wave-numbers.
This property makes it clearly distinct from stochastic
excitation, where the generated frequencies are
similar to the life-times of the source perturbations.
In contrast, frequencies of the oscillations generated by
the mean flow velocity shear induced mode conversion
may be qualitatively higher than the temporal
variation scales of the perturbations in the source flow of a
compressible convection. The frequency spectrum of the excited
acoustic waves should be intrinsically correlated to the velocity
field of the turbulent source flow. Shear flow induced
wave excitation in stratified flows offers a natural
explanation of the fact, that the solar acoustic
oscillation are mainly excited in the high shear regions of the
convection, intergranular dark lanes [29]. It also explains
the puzzling
wave-number dependence of the observed mode energies at fixed
frequencies (see [5] and references therein).
A detailed comparison with observational data
requires a more realistic physical model. The simplicity of
our model is used to demonstrate the basic features of this
excitation phenomenon.

Finally we note that in the present formalism we have focused
on the waves with frequencies higher than the characteristic
cut-off frequency for the acoustic waves in the convection zone.
Shear flow initiates the qualitative change of the
temporal variation scales of perturbations and the excitation of
the waves that are not trapped in the convective envelope.
Hence, this mode conversion presents a new significant
contribution
into the channel of energy transfer from the dynamically active
interior to the atmosphere of the Sun.

~~
\vskip 0.5cm
~~

\section*{ACKNOWLEDGMENTS}
A. G. Tevzadze would like to acknowledge the financial support
as  "bursaal" of the "FWO Vlaanderen", project G.0335.98.
This work was supported in part by the INTAS grant GE97-0504.

\end{document}